\documentclass[hidelinks]{article}
\usepackage{spconf}
\usepackage{cite}
\usepackage{amsmath,amssymb,amsfonts,mathtools}
\DeclareMathOperator*{\argmin}{arg\,min}
\usepackage{algorithm, algpseudocode}
\usepackage{graphicx}
\usepackage{textcomp}
\usepackage{xcolor}
\usepackage{bm}
\usepackage{float}
\usepackage{soul}
\usepackage[acronym,shortcuts]{glossaries}
\usepackage{outlines}
\usepackage[english]{babel}
\usepackage{blindtext}
\usepackage{orcidlink}
\usepackage{subcaption}
\usepackage[]{hyperref}
\usepackage[numbers,sort&compress]{natbib}
\usepackage{lipsum}
\def\BibTeX{{\rm B\kern-.05em{\sc i\kern-.025em b}\kern-.08em
T\kern-.1667em\lower.7ex\hbox{E}\kern-.125emX}}
\usepackage{balance}
\usepackage[bottom]{footmisc}
\usepackage{orcidlink}

\newacronym{OTAC}{OTAC}{over-the-air computing}
\newacronym{AIRCOMP}{AirComp}{over-the-air computing}
\newacronym{ICC}{ICC}{integrated communication and computing}
\newacronym{ISCC}{ISCC}{integrated sensing, communication, and computing}
\newacronym{IOT}{IoT}{internet of things}
\newacronym{QSM}{QSM}{quadrature spatial modulation}
\newacronym{SM}{SM}{spatial modulation}
\newacronym{MMSE}{MMSE}{minimum mean square error}
\newacronym{AP}{AP}{access point}
\newacronym{NMSE}{NMSE}{normalized mean square error}
\newacronym{SIMO}{SIMO}{single-input multiple-output}
\newacronym{B5G}{B5G}{beyond fifth generation}
\newacronym{6G}{6G}{sixth generation}
\newacronym{IQ}{IQ}{in-phase and quadrature}
\newacronym{Q-OTAC}{Q-OTAC}{quadrature over-the-air computing}
\newacronym{SISO}{SISO}{single input single output}
\newacronym{TX}{TX}{transmitter}
\newacronym{RX}{RX}{receiver}
\newacronym{IoT}{IoT}{Internet-of-things}
\newacronym{AI/ML}{AI/ML}{artifitial intelligence/machine learning}
\newacronym{SotA}{SotA}{state-of-the-art}
\newacronym{AWGN}{AWGN}{additive white Gaussian noise}
\newacronym{SNR}{SNR}{signal-to-noise-ratio}
\newacronym{MSE}{MSE}{mean square error}
\newacronym{ED}{ED}{edge device}
\newacronym{CSI}{CSI}{channel state information}
\newacronym{CDF}{CDF}{cumulative distribution function}
\newacronym{ISAC}{ISAC}{integrated sensing and communications}

\title{\vspace{-5ex} Quadrature Over-the-Air-Computing \\ for Multimodal Dual-Stream Signal Processing \vspace{-1ex}}
\name{Hyeon Seok Rou$^{*}$, Kengo Ando$^{*}$,
Giuseppe Thadeu Freitas de Abreu$^{*}$, and David Gonz{\'a}lez G.$^{\dagger}$\vspace{-1ex}}
\address{$^{*}$ School of Computer Science and Engineering, Constructor University, Bremen, Germany \\[0.25ex]
$^{\dagger}$ AUMOVIO Germany GmbH, Frankfurt am Main, Germany \vspace{-2ex}}
\begin{document}
%
\maketitle
\begin{abstract}

We propose a novel \textit{\ac{Q-OTAC}} framework that enables the simultaneously computation of two independent functions and/or data streams within a single transmission.
In contrast to conventional \acs{OTAC} schemes, where a single function is computed by treating each complex signal as a single component, the proposed \ac{Q-OTAC} exploits both \ac{IQ} components of a complex signal, encoding two distinct functions and/or data streams at the \acp{ED} and employing a novel low-complexity \ac{IQ}-decoupled combiner at the \ac{AP} to independently recover each stream, which effectively doubles the computation rate.
A key strength of this framework lies in its simplicity and broad compatibility: the extension into the quadrature domain is conceptually straightforward, yet remakably powerful, allowing seamless integration into existing \acs{OTAC} techniques.
Simulation results validate the effectiveness of this approach, including the first demonstration of dual-function aggregation (e.g., parallel summation and product), highlighting the potential of \ac{Q-OTAC} for enabling multi-modal and high-efficiency \ac{B5G} applications.
\end{abstract}

\vspace{-1.5ex}
\section{Introduction}
\vspace{-1.5ex}

The expected convergence of communication and distributed computing functionalities in \acs{B5G}/\acs{6G} wireless networks \cite{RP-251881, 10588883}, has fueled the development of function-centric communication paradigms, including the increasingly prominent \ac{OTAC} \cite{csahin2023survey,Wang_ITJ24,Liu_TWC20}, also referred to as AirComp.
Rather than transmitting individual data streams for centralized processing, \ac{OTAC} allows multiple edge devices to simultaneously transmit pre-processed signals that are aggregated over the wireless medium -- leveraging the inherent superposition property of the wireless multiple-access channel -- effectively computing a global function directly at the physical layer, via a wide class of nomographic functions, including summations, averages, and geometric means \cite{goldenbaum2014nomographic,Goldenbaum_TSP13}.

Driven by the growing demand for low-latency, energy-efficient, and bandwidth-constrained applications, such as federated learning and sensor fusion in autonomous networks and massive \ac{IOT}, \ac{OTAC} is emerging as a key enabler of \ac{ICC} in future wireless systems \cite{guo2022vehicular,ranasinghe2025flexible}.
Furthermore, its relevance becomes more prominent in a broader context when considering the integration of computation with \ac{ISAC} \cite{liu2022integrated, Rou_SPM2024} -- towards \ac{ISCC} -- where concurrent communications, sensing, and signal processing over wireless channels will be critical to achieve real-time intelligence and system efficiency \cite{Qi_TC22, wen2024survey, li2023integrated}.

However, despite significant research progress - including broadband channel adaptations \cite{qin2021over}, digital combining \cite{liu2024digital}, and correlation handling \cite{frey2021over} - one of the challenges of \ac{OTAC} is essentially related to its computational capacity.
In particular, current methods allow only a single function to be computed per resource instance (e.g., one antenna, one frequency, one symbol).
This limitation arises from the standard practice of mapping data onto the in-phase component of complex wireless signals, while leaving the quadrature component unused for data encoding and only for precoding.
As a result, the system underutilizes the available complex signal space, and the computation rate does not fully harness its potential.

This architectural bottleneck echoes similar limitations that were overcome in other areas of wireless system design - by leveraging the inherent parallel domains of the complex field.
For instance, \ac{QSM} \cite{mesleh2014quadrature} extends conventional \ac{SM} \cite{mesleh2008spatial} by jointly and independently modulating the \ac{IQ} components of transmitted symbols, doubling the spectral efficiency and improving performance without increasing antenna count \cite{rou2022scalable}.

Inspired by such developments, we propose in this article \textit{\acf{Q-OTAC}}, a novel framework that enables the simultaneous computation of \ul{two independent nomographic functions} and/or \ul{two data symbols} by leveraging the \ac{IQ} domains of a single complex-valued transmission.
This is enabled by encoding two streams of pre-processed symbols at the transmitter to each of the \ac{IQ} domains, and applying a novel \ac{IQ}-decoupled \ac{MMSE}-based combiner that independently estimates each target function at the \ac{AP}.

The proposed scheme, to the best of the authors' knowledge, is the first \ac{OTAC}/AirComp method to realize dual-function aggregation per single transmission resource, achieving double the total computation rate.
Moreover, the novel capacity of the \ac{Q-OTAC} to support heterogeneous data inputs and/or functions makes it well-suited for multimodal aggregation tasks, relevant for applications in \ac{B5G}, such as intelligent vehicular networks, edge computing, and semantic inference \cite{clerckx2024multiple, Saad_2020}.


\vspace{-0.5ex}
\section{System Model and Conventional OTAC}
\label{sec:sota}
\vspace{-0.25ex}

Consider a \ac{SIMO} wireless uplink system with $K$ single-antenna \acp{ED} and a single \ac{AP} equipped with $N$ receive antennas.
Under perfect synchronization, the received signal vector $\mathbf{y} \in \mathbb{C}^{N \times 1}$ at the $N$ antennas of the \ac{AP} is given by \vspace{-1ex}
\begin{equation}
\mathbf{y} = \sum_{k=1}^K \mathbf{h}_k p_k s_k + \mathbf{w} = \mathbf{H} \mathbf{P} \mathbf{s} + \mathbf{w} \in \mathbb{C}^{N \times 1},
\label{eq:receivedsignal}
\vspace{-0.5ex}
\end{equation}
where $\mathbf{H} = [\mathbf{h}_1,\ldots,\mathbf{h}_K] \in \mathbb{C}^{N \times K}$ is the channel matrix concatenating the $K$ \ac{ED}-to-\ac{AP} channel vectors $\mathbf{h}_k \in \mathbb{C}^{N \times 1}$, $\mathbf{P} = \mathrm{diag}(\mathbf{p}) \in \mathbb{C}^{K \times K}$ is the diagonal matrix containing the local precoding weights $p_k \in \mathbb{C}$ of the \acp{ED}, collected as a vector in $\mathbf{p} \triangleq [p_1,\ldots,p_K]^\mathsf{T} \in\mathbb{C}^{K \times 1}$, $\mathbf{s} = [s_1,\ldots,s_K]^\mathsf{T} \in \mathbb{C}^{K \times 1}$ is the vector containing the transmit symbols $s_k$ of the $k$-th \ac{ED}, and $\mathbf{w} \sim \mathcal{CN}(\mathbf{0}, \sigma^2_w \mathbf{I}) \in \mathbb{C}^{N \times 1}$ is the \ac{AWGN} with variance $\sigma^2_w \in \mathbb{R}^+$.

In this work, we consider the practical assumption where feedback is unavailable between the \ac{AP} and the \acp{ED}, such that the transmit precoders $p_k$ to be fixed and known a~priori.
This eliminates coordination overhead and improves robustness in dense or latency-sensitive scenarios.
Therefore, in the remainder of this article, we will consider $p_k = 1 ~ \forall k$.


\subsection{State-of-the-Art OTAC Scheme}

In the conventional single-stream \ac{OTAC} setting, the \ac{AP} aims to compute a multivariate target function $f(d_1,\ldots,d_K): \mathbb{R}^K \to \mathbb{R}$ of the data symbols $d_1,\ldots,d_K \in \mathbb{R}$ at the $K$ \acp{ED}.
The target function is assumed to be nomographic\footnotemark[2] \cite{goldenbaum2014nomographic}, which means that it can be decomposed as a post-processed sum of pre-processed symbols, $i.e.$, \vspace{-1.5ex}
\begin{align}
f(d_1,\ldots,d_K) \triangleq \psi\bigg( \sum_{k=1}^K \underbrace{\varphi_k(d_k)}_{\triangleq \; s_k} \bigg) = \psi\bigg( \sum_{k=1}^K s_k \bigg) \in \mathbb{R}, \nonumber \\[-4ex]
\label{eq:OTACtargetfunction}
\end{align}

\vspace{-0.75ex}
\noindent where $\varphi_k(\cdot)$ denotes the local pre-processing function at the $k$-th \ac{ED}, $\psi(\cdot)$ denotes the global post-processing function at the \ac{AP}, and $s_k \triangleq \varphi_k(d_k) \in \mathbb{R}$ has been defined as the effective transmit symbol of the $k$-th \ac{ED} after pre-processing - and is ultimately equivalent to the transmit symbols of eq. \eqref{eq:receivedsignal}.

For example, when the target function is a summation, the pre-/post-processing functions are simply identity functions, $i.e.$, $\varphi_k^{\mathrm{SUM}}(x) = x$ and $\psi^{\mathrm{SUM}}(x) = x$, such that $f^\mathrm{SUM}(d_1,\ldots,d_K) \!\triangleq\! \psi^{\mathrm{SUM}} \big( \sum_{k=1}^K \varphi_k^{\mathrm{SUM}}(d_k)\big) \!=\! \sum_{k=1}^{K} d_k.$

\setcounter{footnote}{1}
\footnotetext{Many real-valued multivariable functions used in practice -- such as summation, average, weighted average, product, geometric mean, $p$-norms, and polynomials -- are directly nomographic and thus well-suited for \ac{OTAC}.}

On the other hand, when the target function is a product, the pre-/post-processing functions are the logarithmic and exponentiation to an arbitrary base $b$, $i.e.$, $\varphi_k^{\mathrm{PROD}}(x) = \log_b(x)$ and $\psi^{\mathrm{PROD}}(x) = b^{\;\!x} = \mathrm{exp}_b(x)$, respectively, such that $f^\mathrm{PROD}(d_1,\ldots,d_K) \triangleq \exp_b\!\big( \sum_{k=1}^K \log_b(d_k) \big) \! = \! \prod_{k=1}^{K} d_k.$
%

Then, given that the transmit symbols at the \acp{ED} are generated via pre-processing the data symbols as $s_k \triangleq \varphi_k(d_k) \in \mathbb{R}$, the received signal model of eq. \eqref{eq:receivedsignal} can be rewritten as \vspace{-1ex}
\begin{equation}
\mathbf{y} = \sum_{k=1}^K \mathbf{h}_k s_k + \mathbf{w} = \sum_{k=1}^K \mathbf{h}_k \varphi_k(d_k) + \mathbf{w} \in \mathbb{C}^{N \times 1},
\label{eq:receivedsignal_OTAC}
\vspace{-1ex}
\end{equation}
where it can be seen that the wireless multiple-access channel inherently achieves the summation of the pre-processed symbols via the  superposition phenomena.

Therefore, the objective of the \ac{AP} is to remove the effect of the channel to yield the summation of only the pre-processed symbols, such that the post-processing function can be applied to find the target function output.
Of course, each symbol can be estimated independently then recombined to yield the desired sum, but it should be stated clearly that this is \underline{not} the objective of \ac{OTAC}, and instead, the objective is to apply a single combiner vector $\mathbf{u} \in \mathbb{C}^{N \times 1}$ to the received signal to obtain the estimate of the target function, $i.e.$, \vspace{-1ex}
\begin{equation}
\tilde{f}(\mathbf{u}; \mathbf{y}) = \psi\big(\mathbf{u}^{\mathsf{T}} \mathbf{y}\big) \approx f(d_1,\ldots,d_K) \in \mathbb{R}.
\label{eq:OTACcombinerbased_targetfunction}
\end{equation}

In all, given the channel matrix $\mathbf{H}$, the \ac{AP} aims to construct an optimal combiner which minimizes the \ac{MSE} of the estimated target function\footnote{Assuming a feedback channel between the \ac{AP} and the \acp{ED}, the precoding weights $p_k$ can also be jointly optimized with the combiner $\mathbf{u}$ in order to minimize the objective function.
However, this consideration is out of scope of this article, and will be considered in an extended work.}, formulated as \vspace{-1ex}
\begin{align}
\mathbf{u}_\mathrm{opt} & = \argmin_{\mathbf{u}} \,\mathbb{E}\!\left[\left| f(d_1,\ldots,d_K) - \psi(\mathbf{u}^{\mathsf{T}} \mathbf{y}) \right|^2\right].
\label{eq:MSE_single_stream}
\vspace{-2ex}
\end{align}

\vspace{-1ex}
While various \ac{SotA} methods exist for designing $\mathbf{u}$ -- such as branch-and-bound, iterative, and message-passing schemes -- we focus on \ac{MMSE}-based linear combiners due to their simplicity in admitting a closed-form solution with a strong baseline performance \cite{ando2025bayesian,Ando_CAMSAP23}, with \vspace{-1ex}
\begin{equation}
\mathbf{u}_\mathrm{MMSE} = \big(\mathbf{H} \mathbf{H}^\mathsf{H} + \sigma^2_w\mathbf{I}_{N\times N} \big)^{\!-1} \!\cdot  \mathbf{H} \cdot \mathbf{1}_{K \times 1} \in \mathbb{C}^{N \times 1}.
\label{eq:MMSE_combiner_sota} \vspace{-1ex}
\end{equation}

Since the combiner in eq. \eqref{eq:MMSE_combiner_sota} is complex-valued, the combined output $\mathbf{u}^{\mathsf{T}} \mathbf{y}$ is also complex.
However, since the original data symbols and the nomographic target functions in \ac{OTAC} are real-valued, as defined in eq.~\eqref{eq:OTACtargetfunction}, only the real part of the combined signal is relevant.
Thus, in practice, the function estimate is obtained using the post-processing function only unto the real component of the combined output, $i.e.$, \vspace{-1ex}
\begin{equation}
\tilde{f}(\mathbf{u}; \mathbf{y}) = \psi\!\left(\mathrm{Re}\{\mathbf{u}^{\mathsf{T}} \mathbf{y}\}\right) \in \mathbb{R}.
\label{eq:MMSE_combined_postprocessing}
\vspace{0.75ex}
\end{equation}

\vspace{-2ex}
\section{Proposed Dual-Stream Q-OTAC}
\label{sec:dualstream_systemmodel}
\vspace{-1ex}

As can be seen in Section \ref{sec:sota}, the conventional \ac{OTAC} paradigm processes only a single real-valued function of scalar data symbols per transmission, fundamentally underutiliing the complex signal space by discarding the quadrature component, and also inherently limiting the computational capacity of each resource instance to one computing stream.

\begin{figure}[t]
\centering
\includegraphics[width=1\columnwidth]{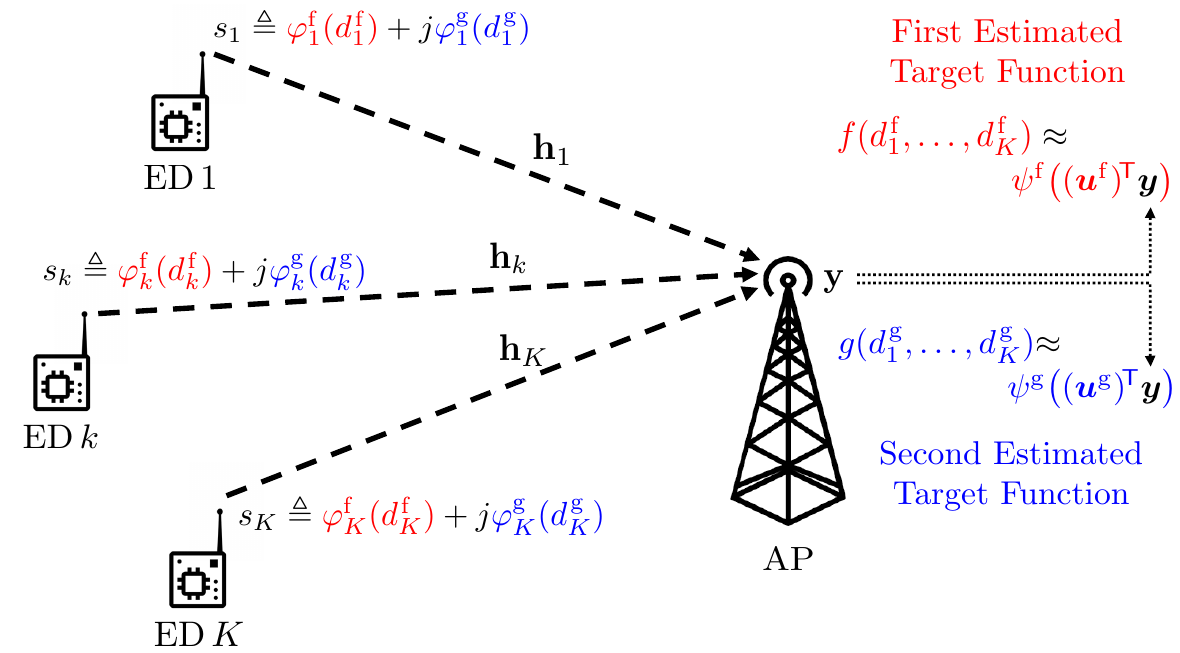}
\vspace{-3ex}
\caption{Illustration of the proposed \ac{Q-OTAC} system.}
\label{fig:scenario_proposed}
\vspace{-2ex}
\end{figure}

To address this limitation, we propose a dual-stream \acf{Q-OTAC} framework that encodes two real-valued data symbols -- one in each of the \ac{IQ} components of the complex transmit symbol -- with a novel combiner design which enables an independent computation of the two nomographic functions at the \ac{AP}, which has been illustrated in Fig.~\ref{fig:scenario_proposed}, clearly highlighting the two streams.

\vspace{-1.5ex}
\subsection{Proposed Q-OTAC Dual-stream  Transmitter Design}
\label{sec:dualstream_transmitter}

First, let us denote the two independent \ac{OTAC} target functions by $f(\,\cdots) : \mathbb{R}^K \to \mathbb{R}$ and $g(\,\cdots) : \mathbb{R}^K \to \mathbb{R}$, with the corresponding pre-/post-processing functions given by $\psi^\mathrm{f}(\cdot), \psi^\mathrm{g}(\cdot)$ and $\varphi^\mathrm{f}_k(\cdot), \varphi^\mathrm{g}_k(\cdot)$, respectively, such that
\begin{subequations}
\begin{align}
f(d^\mathrm{\;\!f}_1,\ldots,d^\mathrm{\;\!f}_K) & \triangleq \psi^\mathrm{f}\Big( \textstyle\sum_{k=1}^K \varphi^\mathrm{f}_k(d^\mathrm{\;\!f}_k)\Big)\in \mathbb{R},  \\[1ex]
g(d^\mathrm{\;\!g}_1,\ldots,d^\mathrm{\;\!g}_K) & \triangleq \psi^\mathrm{g}\Big( \textstyle\sum_{k=1}^K \varphi^\mathrm{g}_k(d^\mathrm{\;\!g}_k)\Big)\in \mathbb{R}, 
\label{eq:QOTACtargetfunction}
\vspace{-0.5ex}
\end{align}
\vspace{-2.5ex}

\noindent where $d^\mathrm{\;\!f}_k \in \mathbb{R}$ and $d^\mathrm{\;\!g}_k \in \mathbb{R}$ are the two independent data stream at the $k$-th \ac{ED}, respectively for each target function.
\end{subequations}

In light of the above, the \ac{Q-OTAC} scheme proposes to transmit the two streams of the pre-processed data symbols $s^\mathrm{f}_k \!\triangleq\! \varphi_k^\mathrm{f}(d_k^\mathrm{\;\!f}) \!\in\! \mathbb{R}$ and $s^\mathrm{g}_k \!\triangleq\! \varphi_k^\mathrm{g}(d_k^\mathrm{\;\!g}) \!\in\! \mathbb{R}$ respectively in the \ac{IQ} domain of the complex transmit symbol, $i.e.$, \vspace{-0.5ex}
\begin{equation}
s_k \triangleq s^\mathrm{f}_k  + js^\mathrm{g}_k =  \varphi_k^\mathrm{f}(d_k^\mathrm{\;\!f}) + j\varphi_k^\mathrm{g}(d_k^\mathrm{\;\!g}) \in \mathbb{C}, \vspace{-0.5ex}
\label{eq:QOTAC_tx_symbols}
\end{equation}
where $j \triangleq \sqrt{-1}$ is the imaginary unit.

The resulting received signal, with $p_k = 1$, is given by \vspace{-1ex}
\begin{equation}
\mathbf{y} = \sum_{k=1}^{K} \mathbf{h}_k \big(s_k^\mathrm{f} \!+\! j s_k^\mathrm{g}\big) + \mathbf{w} = \mathbf{H} (\mathbf{s}^\mathrm{f} \!+\! j\mathbf{s}^{\mathrm{g}}) + \mathbf{w} \!\in\! \mathbb{C}^{N \times 1} \label{eq:receivedsignal_QTOAC}
\vspace{-1,5ex}
\end{equation}
where $\mathbf{s}^{\mathrm{f}} \!\triangleq\! [s^{\mathrm{f}}_1,\ldots,s^{\mathrm{f}}_{K}]^{\mathsf{T}} \!\in\! \mathbb{R}^{N \times 1}$, $\mathbf{s}^{\mathrm{g}} \!\triangleq\! [s^{\mathrm{g}}_1,\ldots,s^{\mathrm{g}}_{K}]^{\mathsf{T}} \!\in\! \mathbb{R}^{N \times 1}$ are the vectors containing the in-phase and quadrature symbols transmitted by the \acp{ED}, respectively.

Equation \eqref{eq:receivedsignal_QTOAC} illustrates the key novelty of the proposed \ac{Q-OTAC} framework in the complex-valued transmit symbol containing two data streams in each of the \ac{IQ} components, contrasting with the conventional \ac{OTAC} in eq.~\eqref{eq:receivedsignal_OTAC} with only a real-valued transmit symbol composed of only one stream.

\vspace{-1.5ex}
\subsection{Proposed Q-OTAC Dual-stream Combiner Design}
\label{sec:dualstream_combiner}

First, the received signal model of eq. \eqref{eq:receivedsignal_QTOAC} is decoupled and reformulated to a real-valued equivalent system as
\begin{align}
\bm{y} \triangleq 
\begin{bmatrix}
\Re\{ \mathbf{y} \}\\
\Im\{ \mathbf{y} \}
\end{bmatrix}
\!=\! 
\underbrace{
\begin{bmatrix}
\Re\{ \mathbf{H}\} &\hspace{-1.5ex} -\Im\{ \mathbf{H} \} \\
\Im\{ \mathbf{H} \} & \Re\{ \mathbf{H} \}
\end{bmatrix}}_{\triangleq \bm{H} \in \mathbb{R}^{2N \times 2N}}
\hspace{-2.25ex}
\overbrace{\begin{bmatrix}
\mathbf{s}^\mathrm{f}\\
\mathbf{s}^\mathrm{g}\\
\end{bmatrix}}^{\triangleq \bm{s} \in \mathbb{R}^{2N \times 1}}
\hspace{-2.25ex}+\! \!
\underbrace{\begin{bmatrix}
\Re\{ \mathbf{w} \}\\
\Im\{ \mathbf{w} \}
\end{bmatrix}}_{\triangleq \bm{w} \in \mathbb{R}^{2N \times 1}}
\! \!\in \mathbb{R}^{2N \times 1},
\nonumber \\[-3.5ex]
\label{eq:receivedsignal_dualstream_decoupled} 
\end{align}
where $\bm{y} \in \mathbb{R}^{2N \times 1}, \bm{H} \in \mathbb{R}^{2N \times 2N}, \bm{s} \in \mathbb{R}^{2N \times 1}, \bm{w} \in \mathbb{R}^{2N \times 1}$ are the \ac{IQ}-decoupled received signal vector, transmit signal vector and \ac{AWGN} vector, respectively.

In light of the above, we aim to design two distinct combiners $\bm{u}^\mathrm{f} \in \mathbb{R}^{2N \times 1}$ and $\bm{u}^\mathrm{g} \in \mathbb{R}^{2N \times 1}$ to be applied directly unto the received signal vector ${\bm{y}}$, which respectively yields the estimated target function values $\tilde{f}$ and $\tilde{g}$, $i.e.$,
\begin{subequations}
\begin{align}
\tilde{{f}}(\bm{u}^\mathrm{f};\bm{y}) &= \psi^\mathrm{f}\big((\bm{u}^\mathrm{f}){}^\mathsf{\!T}\,\!\bm{y}\big) \,\approx f(d^\mathrm{\;\!f}_1,\ldots,d^\mathrm{\;\!f}_K) \in \mathbb{R},
\label{eq:estimated_targetfunction_I} \\
\tilde{g}(\bm{u}^\mathrm{g};\bm{y}) &= \psi^\mathrm{g}\big((\bm{u}^\mathrm{g}){}^\mathsf{\!T}\,\!\bm{y}\big) \approx g(d^\mathrm{\;\!g}_1,\ldots,d^\mathrm{\;\!g}_K) \in \mathbb{R}.
\label{eq:estimated_targetfunction_Q}
\end{align}
\label{eq:estimated_Qtargetfunctions}
\vspace{-2.5ex}
\end{subequations}

Given the above, the \ac{MMSE}\footnote{As with conventional \ac{OTAC}, other optimization methods for combiner design may be considered, however, in this work, we focus on the closed-form, low-complexity linear \ac{MMSE}-based  combiner as a proof of concept of the proposed dual-stream framework.} problem is formulated as
\begin{subequations}
\begin{align}
\mathbf{u}^{\mathrm{f}}_\mathrm{opt} & = \argmin_{\mathbf{u}^\mathrm{f}} \,\mathbb{E}\Big[\big| f(d^\mathrm{\;\!f}_1,\ldots,d^\mathrm{\;\!f}_K) - \tilde{f}(\bm{u}^\mathrm{f};\bm{y}) \big|^2\Big]
\label{eq:optu_f} \\[-1ex]
& \equiv \argmin_{\mathbf{u}^\mathrm{f}} \,\mathbb{E}\Big[\big| \underbrace{\mathbf{1}_{N\times 1}^{\mathsf{T}}\!\cdot\!\mathbf{s}^\mathrm{f}}_{\sum_{k=1}^{K} \!s^{\mathrm{f}}_k} \, - \,(\bm{u}^\mathrm{f})^\mathsf{\!T} \,\!\!\cdot\! (\bm{H}\bm{s} + \bm{w}) \big|^2\Big] \nonumber, \\[0.5ex]
\mathbf{u}^{\mathrm{g}}_\mathrm{opt} & = \argmin_{\mathbf{u}^\mathrm{g}} \,\mathbb{E}\Big[\big| g(d^\mathrm{\;\!g}_1,\ldots,d^\mathrm{\;\!g}_K) - \tilde{g}(\bm{u}^\mathrm{g};\bm{y}) \big|^2\Big]
\label{eq:optu_g} \\[-1ex]
& \equiv \argmin_{\mathbf{u}^\mathrm{g}} \,\mathbb{E}\Big[\big| \underbrace{\mathbf{1}_{N\times 1}^{\mathsf{T}}\!\cdot\!\mathbf{s}^\mathrm{g}}_{\sum_{k=1}^{K} \!s^{\mathrm{g}}_k} \, - \,(\bm{u}^\mathrm{g})^\mathsf{\!T} \,\!\!\cdot\! (\bm{H}\bm{s} + \bm{w}) \big|^2\Big] \nonumber,
\end{align}
\label{eq:QOTAC_MMSE_min}
\vspace{-1ex}
\end{subequations}

\noindent where the respective inverses of the post-processing functions has been applied to the two terms of the squared error, while retaining the equivalence of the minimization problems as the post-processing function of nomographic functions are typically strictly monotonic ($i.e.$, identity, scalar division, exponentiation) such that the minimization of the squared error in the transformed domain, $i.e.$, $|\psi(b) - \psi(a)|^2$, is equivalent to the minimization in the original domain, $|b - a|^2$.

In light of the above, the closed-form solutions of the \ac{MMSE} combiners of eq. \eqref{eq:QOTAC_MMSE_min} are efficiently obtained as
\begin{subequations}
\begin{align}
\bm{u}_\mathrm{MMSE}^{\mathrm{f}} &= \! \big(({\bm{H}} {\bm{H}}^\mathsf{T} \!+\!\tfrac{\sigma^2_w}{2}\mathbf{I}_{2N\times2N}) \big)^{\!-1} {\bm{H}}\bm{c}^\mathrm{f} \in \mathbb{R}^{2N \times 1}\!,\!\!\\
\bm{u}_\mathrm{MMSE}^{\mathrm{g}} &= \! \big(({\bm{H}} {\bm{H}}^\mathsf{T} \!+\! \tfrac{\sigma^2_w}{2}\mathbf{I}_{2N\times2N}) \big)^{\!-1} {\bm{H}}\bm{c}^\mathrm{g}\!\in \mathbb{R}^{2N \times 1}\!,\!\!
\end{align}
\end{subequations}
where the auxiliary domain selector vectors $\bm{c}^\mathrm{f}  \in \mathbb{R}^{2N \times 1}$ and $\bm{c}^\mathrm{g}  \in \mathbb{R}^{2N \times 1}$ are simply given by
\begin{equation}
\bm{c}^\mathrm{f} \triangleq 
\begin{bmatrix}
\mathbf{1}_{N \times 1} \\
\mathbf{0}_{N \times 1}
\end{bmatrix} \in \mathbb{R}^{2N \times 1}, ~\text{and}~
\bm{c}^\mathrm{g} \triangleq 
\begin{bmatrix}
\mathbf{0}_{N \times 1} \\
\mathbf{1}_{N \times 1}
\end{bmatrix} \in \mathbb{R}^{2N \times 1}.
\end{equation}

In all, by configuring the pre/post-processing functions of two independent target functions, the proposed \ac{Q-OTAC} framework enables the parallel computation of two distinct target functions and/or of two distinct sets of data ($i.e.$, two independent streams), by applying two highly efficient \ac{MMSE}-based combiners $\bm{c}^\mathrm{f}$ and $\bm{c}^\mathrm{g}$ to the received signal.

\vspace{-2ex}
\section{Performance Analysis}
\label{sec:performanceanalysis}
\vspace{-1.5ex}

\begin{figure}[b!]
\centering
\vspace{-2ex}
\includegraphics[width=0.9\columnwidth]{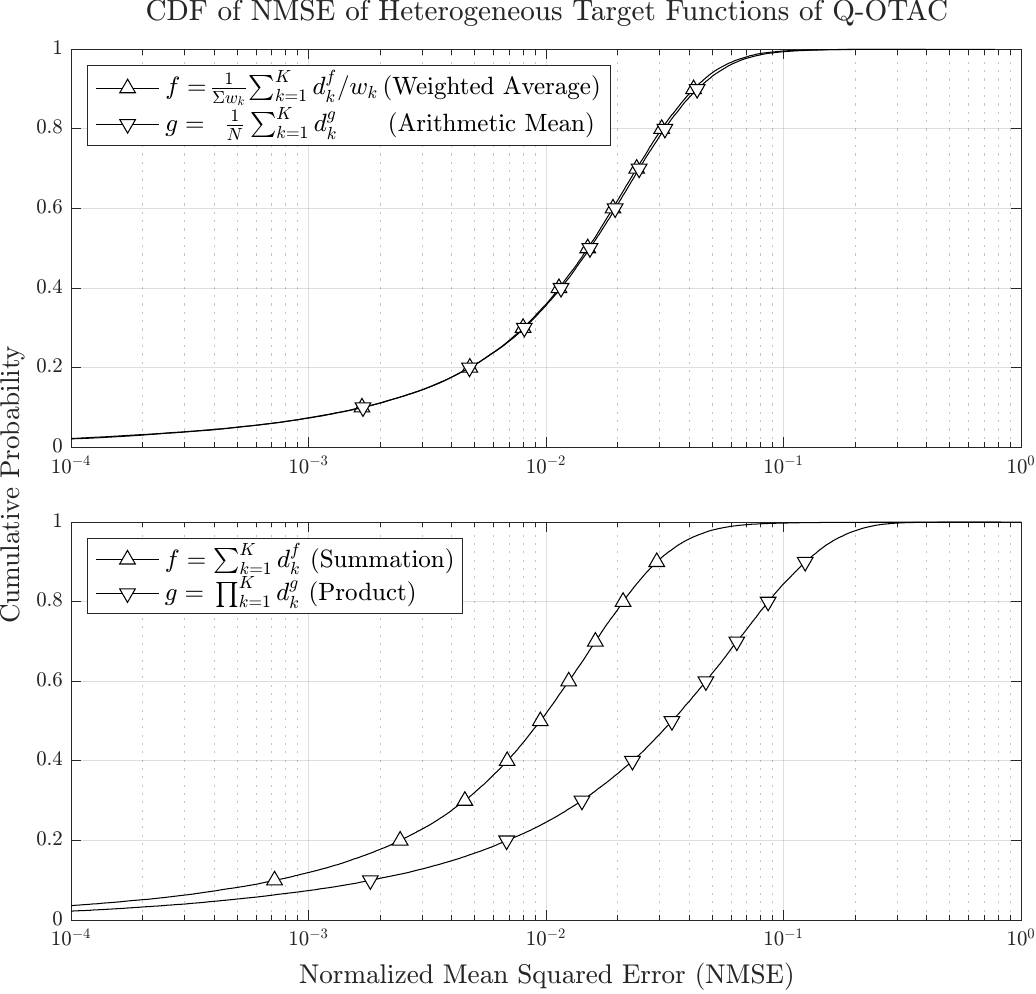}
\vspace{-1ex}
\caption{The \acs{NMSE} \acs{CDF} for different multimodal dual-stream \ac{Q-OTAC} scenarios for $N = 20, K = 20$ and $\mathrm{SNR} = 15\mathrm{dB}$.}
\label{fig:Multifunction_NMSE_CDF}
\end{figure}

In this section, we demonstrate the core innovation of the proposed dual-stream \ac{Q-OTAC} scheme and provide the first proof-of-concept that the \ac{OTAC} framework can achieve simultaneous computation of two distinct target functions using only a single wireless transmission resource and a low-complexity (closed-form) linear combiner at the \ac{AP}.
Specifically, the performance of the proposed dual-stream \ac{Q-OTAC} scheme is validated and analyzed via numerical simulation, comparing against the conventional single-stream \ac{OTAC} baseline using the linear \ac{MMSE} combiner of eq. \eqref{eq:MMSE_combiner_sota}.

First, Figure~\ref{fig:Multifunction_NMSE_CDF} presents the \acp{CDF} of the \ac{NMSE} for various heterogeneous function pairs, including simultaneous arithmetic mean and weighted average (with arbitrary weights), as well as a simultaneous summation and product. 
In all cases, the proposed \ac{Q-OTAC} scheme maintains reliable estimation accuracy across functions.

A relative performance variation among the different function combinations can also be observed, but this is not a remnant of the combiner design, but reflects the underlying numerical behavior of the pre-/post-processing of the nomographic functions.
For example, the product function involves logarithmic and exponential operations, which are inherently more sensitive to noise than the identity mappings used for the summation and mean.

\begin{figure}[t]
\centering
\begin{subfigure}{0.925\columnwidth}
\centering
\includegraphics[width=\linewidth]{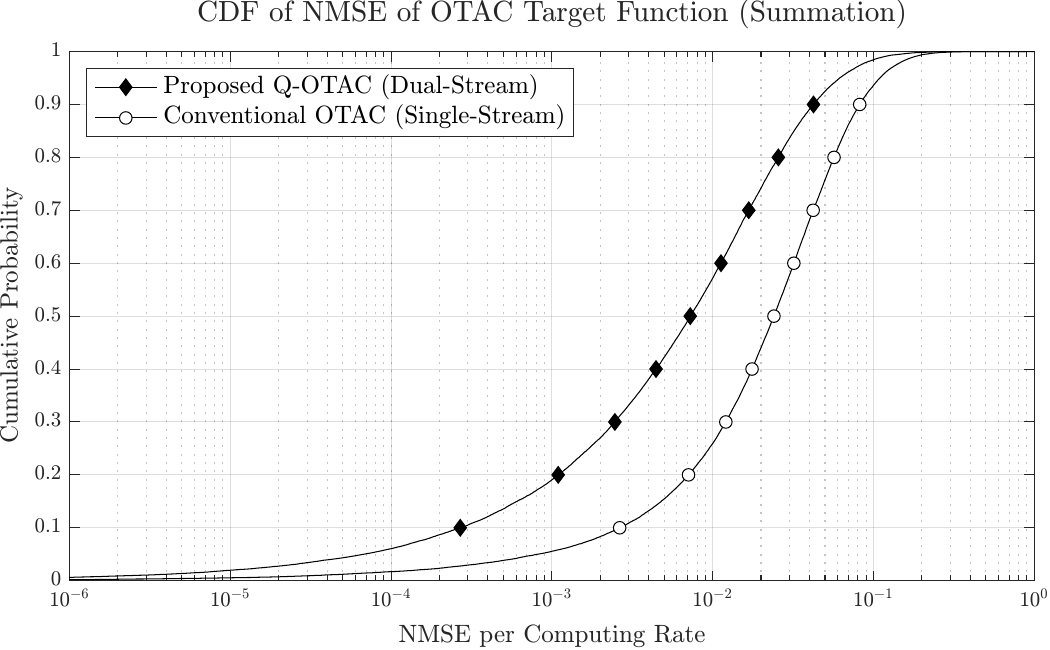}
\caption{\Ac{CDF} of the \ac{NMSE} ($\mathrm{SNR = 15\text{dB}}$).}
\label{fig:NMSE_CDF}
\end{subfigure}
\vspace{1.5ex}

\noindent \hspace{-2.5ex} \begin{subfigure}{0.925\columnwidth}
\centering
\includegraphics[width=\linewidth]{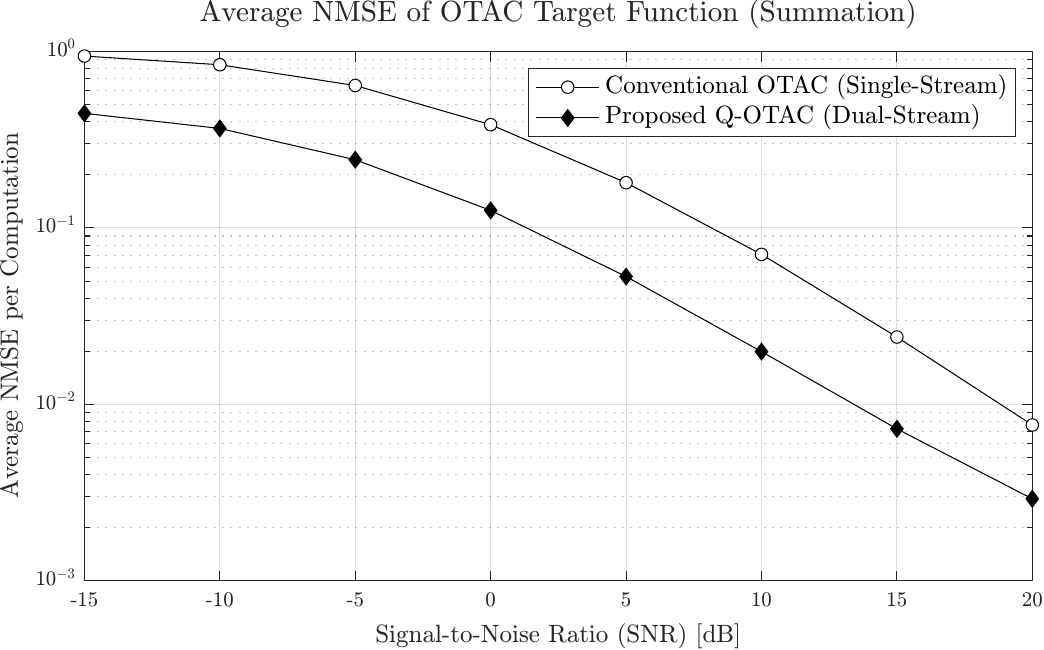}
\caption{Average \ac{NMSE} against \acs{SNR}.}
\label{fig:ANMSE_SNR}
\end{subfigure}
\vspace{-1ex}
\caption{Summation performance of the proposed \ac{Q-OTAC} against the conventional \ac{OTAC}, with $N = 20$, $K = 20$.}
\label{fig:QOTAC_performance}
\vspace{-3ex}
\end{figure}

Next, Figure~\ref{fig:QOTAC_performance} compares the performance of the proposed dual-stream \ac{Q-OTAC} with the conventional single-stream \ac{OTAC} scheme, evaluated in terms of the total \ac{NMSE} per computation via the \ac{CDF} and average \ac{NMSE} over \acs{SNR}, in the respective subfigures.
The results demonstrate that \ac{Q-OTAC} achieves a noticable improved total \ac{NMSE} per computation, over the entire range of both the \ac{CDF} and \acs{SNR} plots with an approximate gain of $~5\text{dB}$ thanks to its novel and efficient utilization of both the \ac{IQ} components of the complex signal to perform computing, as opposed to only a single real component in the \ac{SotA}, thereby doubling the signal space and effective computation rate.

\vspace{-1ex}
\section{Conclusion and Future Works}
\vspace{-1ex}

We proposed the \textit{\acf{Q-OTAC}}, the first \ac{OTAC} framework to enable dual-function aggregation in a single transmission via a novel \ac{IQ}-decoupled \ac{MMSE} combiner.
Simulation results confirm its effectiveness in heterogeneous function settings, achieving twice the computation rate and improved average accuracy over the single-stream \ac{OTAC}, supporting multimodal and scalable \ac{B5G} \ac{OTAC} scenarios.
Future works will investigate extended combiner designs, jointly with the precoder optimization considerations, and incoporate \ac{ICC}/\ac{ISCC} approaches.



\end{document}